# A Study of Temperature-dependent Properties of N-type δ-doped Si Band-structures in Equilibrium

Hoon Ryu*, Sunhee Lee and Gerhard Klimeck
Network for Computational Nanotechnology, Purdue University, West Lafayette, IN 47907, USA
*e-mail: ryu2@purdue.edu

*Abstract* — **A highly phosphorus δ-doped Si device is modeled with a quantum well with periodic boundary conditions and the semi-empirical spds* tight-binding band model. Its temperature-dependent electronic properties are studied. To account for high doping density with many electrons, a highly parallelized self-consistent Schrödinger-Poisson solver is used with atomistic representations of multiple impurity ions. The band-structure in equilibrium and the corresponding Fermi-level position are computed for a selective set of temperatures. The result at room temperature is compared with previous studies and the temperature-dependent electronic properties are discussed further in detail with the calculated 3-D self-consistent potential profile.**

Keywords: δ-doped Si; Schrödinger-Poisson; Self-consistent simulation; Tight-binding;

## I. INTRODUCTION

### A. Need for δ-doped Si Devices

δ-doped Si devices with high phosphorous doping (Si:P) have attracted the attention of experimentalists due to their potential utility in quantum computing application as well as in reservoirs of electrons for quantum-wires [1, 2]. Theoretical understanding of electronic properties in such systems based on a realistic modeling approach is critical in potential device designs. Three recent studies have demonstrated the electronic property of Si:P devices in equilibrium. They were, however, limited in the assumption of uniform-doping density [3], only 1-D variation of charge-potential profiles in describing δ-doped layers [4] and a periodic stack of δ-doped layers [5].

### B. Goals of the work

The objective of this work is to provide a theoretical study for temperature-dependent electronic properties of Si:P devices through a full 3-D self-consistent simulation using the semi-empirical sp3d5s* tight-binding band model without spin-orbit coupling. The tight-binding approach based on the atomistic representation will enable us to overcome the limitations in the previous works [3-5] and explain semiconductor devices with realistically large dimensions.

## II. METHODOLOGIES

### A. Atomistic Tight-Binding Band Models.

We have recently begun the construction of the new OMEN -3D tool which is based on the well-established NEMO-3D work. [6, 7]. NEMO 3-D can calculate strain and electronic structure for realistically sized systems as large as 52 million atoms which corresponds to a simulation domain of $(101nm)^3$. Strain is calculated using the atomistic Valence Force Field (VFF) method [8] and the electronic structure using a twenty-band sp3d5s* nearest neighbor empirical tight-binding model [9]. The tight-binding parameters are fit to reproduce the bulk properties of GaAs, InAs, AlAs, Si, and Ge with respect to room temperature band edges, effective masses, hydrostatic and biaxial strain behaviors [9-12] using a global minimization procedure based on a genetic algorithm [13] and an analytical insight [10]. For realistic semi-conducting nano-scale systems, our tight binding approach has been validated experimentally by modeling the InGaAs/InAlAs resonant tunneling diode at high bias and high current [14], the photoluminescence in InAs nano-particles [15], the valley-splitting in miscut Si quantum wells on SiGe substrates [16], the gate-induced Stark effect [17] and the transition of quantum-confinements of a single phosphorus impurity in Si FinFET [18, 19]. While previous impurity works focused on a single, isolated and ionized donor ion [17-19], we expend significantly on this scope with charge-potential self-consistent calculations of many impurities in a δ-doped layer.

### B. Charge-Potential Self-Consistent Calculations

The process for self-consistent simulation used in this work, is summarized in Fig. 1. Due to the extremely high doping density in the δ-doped layer, the QW structure should have many electrons to satisfy the charge neutrality therefore a charge-potential self-consistent calculation becomes one of the most important factors to be considered. For this purpose, we have developed a Schrödinger-Poisson solver which is highly parallelized with MPI/C++ and the 3-D Schrödinger equation is solved with the LANCZOS algorithm-based eigen-solver to obtain band-structures and corresponding local charge profiles in equilibrium. 3-D spatial potential distributions are calculated with a Poisson solver which discretizes a spatial domain of devices with the Finite Different Method and uses the Aztec package as a parallel iterative linear solver [20]. Detailed information regarding the parallelized solver and its numerical performance can be found in the reference [21].

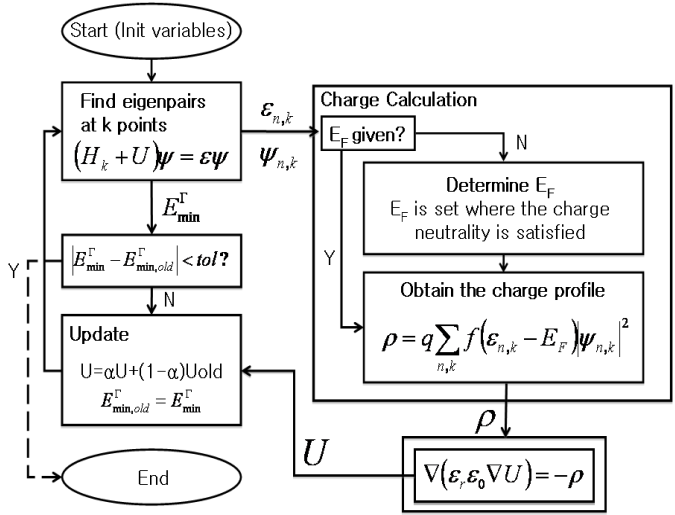

Figure 1. A Flow-chart for the Self-consistent calculation.

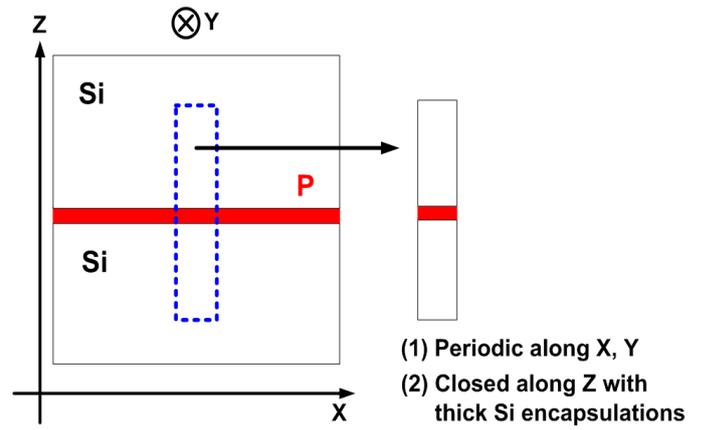

Figure 2. A simplified schematic of the δ-doped Si devices and the quantum well which has been taken for numerically efficient modeling.

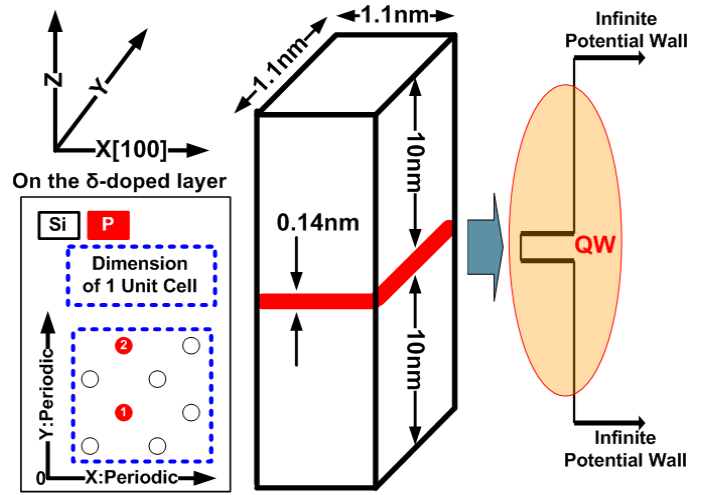

Figure 3. A 3-D atomistic structure of the Si:P quantum well device

## C. Modeling of Device Geometry

δ-doped Si devices are fabricated with following processes [22]: (1) Create a phosphorus sheet on top of Si by annealing a cleaned surface of Si bulk with Phosphine ($PH_3$) gas at high temperature ~ 550(C°), (2) Overgrow the sheet depositing an epitaxial Si film at relatively low temperature ~ 250(C°). This results in an extended sheet of a δ-doped layer embedded in Si bulk. As a first trial for the atomistic study, we focus on an ideal, well-ordered system by taking a small slice from the device with periodic boundary conditions. This preliminary approach ignores the effect of dopant-disorder. It, however, not only reduces computing expense significantly, but allows us to test the ideal physics and explore the capability of our new simulator. The target device for modeling, therefore, becomes a Si:P quantum well (QW) as shown in Fig. 2, where a simplified visualization of the δ-doped Si device is depicted. Note that a dashed-line is used to show the central part of the simulation domain.

Fig. 3 depicts a more elaborate geometry structure of Si:P (QW) which has a strong quantum confinement due to the impurity ion induced potential A thickness of one [100] Si atomic layer for the δ-doped layer and $1.7 \times 10^{14}$ ($cm^{-2}$) for the 2-D sheet doping density is assumed. The particular value corresponds to 2 phosphorus atoms per each atomic layer of $(1.1(nm))^2$ (2×2 [100] Si unit cells) and is chosen to compare our result at room temperature with the previous results [3, 5]. While they assumed a constant doping in the δ-doped layer [3] and a cubic-grid system [5], a realistic doping is considered with an atomistic representation of the impurities based on the zinc-blende crystal. A closed boundary condition is assumed along the z-direction with a thick 20-nm Si encapsulation to minimize artifacts stemming from the infinite potential walls.

## III. RESULTS AND DISCUSSION

### A. Equilibrium Electronic Properties in at room temperature

1-D cuts of the 3-D self-consistent potential profile of the Si:P QW at room temperature are shown in Fig. 4, where the core potential has a minimum of -2.8 (V) at the impurity positions. The work with the sp3d5s* tight binding approach by Rahman et al., [17], has studied a single, isolated and ionized phosphorus atom in Si bulk to require a core correction of -4.33(V) as the potential minimum to match the ground-state energy to experimental results. Compared to the potential minimum of -4.33(V) reported for a single impurity, the self-consistent result here is larger because phosphorus ions are under a strong screening by electrons due to extremely high doping density in the δ-doped layer.

Fig. 5 shows the equilibrium band-structures along [100], [110] directions at room temperature which are computed with the potential profile given in Fig. 4. Here, a zero-energy is used as a reference point to represent the conduction band minima (CBM) of bulk-Si. Due to extremely high doping density in the δ-doped layer, impurity atoms are placed closely enough to create donor sub-bands (donor-levels) below CBM of bulk-Si. At room temperature, the donor-band minimum turns out to be placed at -367 (meV) with a Fermi-level at -87 (meV). Note that previous studies with the same phosphorus doping in δ-doped layer predicted the Fermi-level position at -99 (meV) [3, 5]. The first 4 sub-bands of the Si:P QW 2-D band-structures are plotted upto 5×kT from the Fermi-energy level in Fig. 6, which clearly shows that most of the local electrons in the Si:P QW should come from these donor-levels.

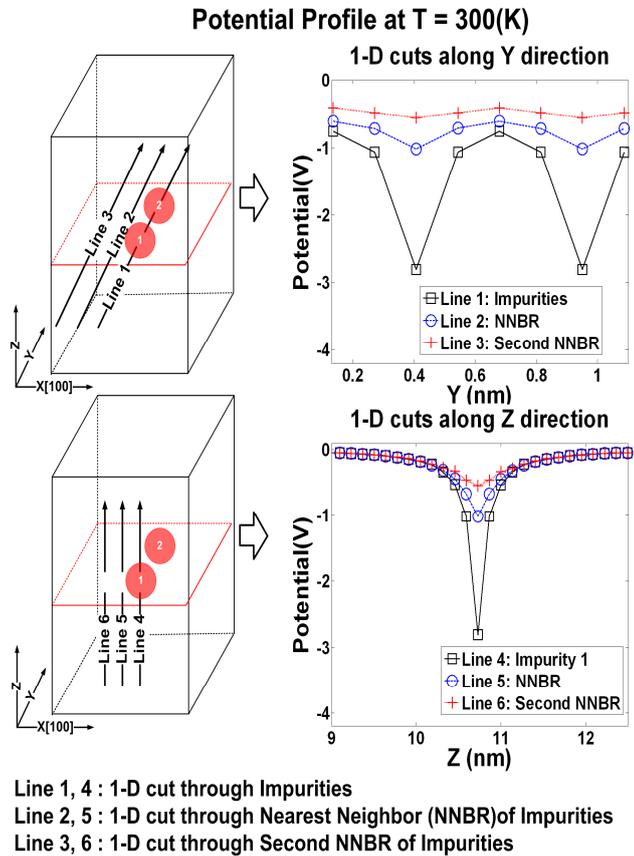

Line 1, 4 : 1-D cut through Impurities
Line 2, 5 : 1-D cut through Nearest Neighbor (NNBR) of Impurities
Line 3, 6 : 1-D cut through Second NNBR of Impurities

Figure 4. Self-consistently calculated potential profiles at room temperature

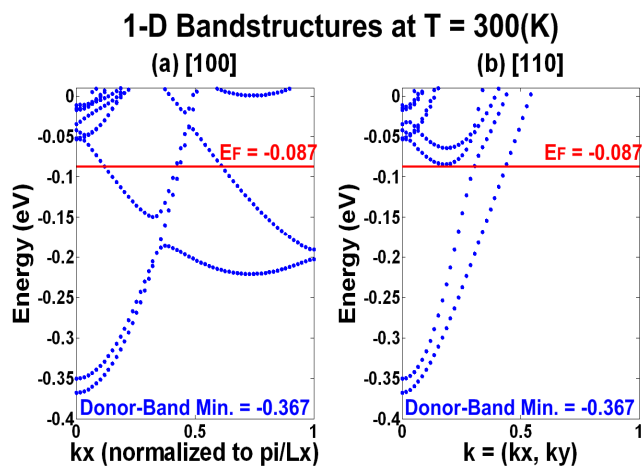

Figure 5. (a) The equilibrium [100] and (b) [110] band-structure of the Si:P quantum well at room temperature

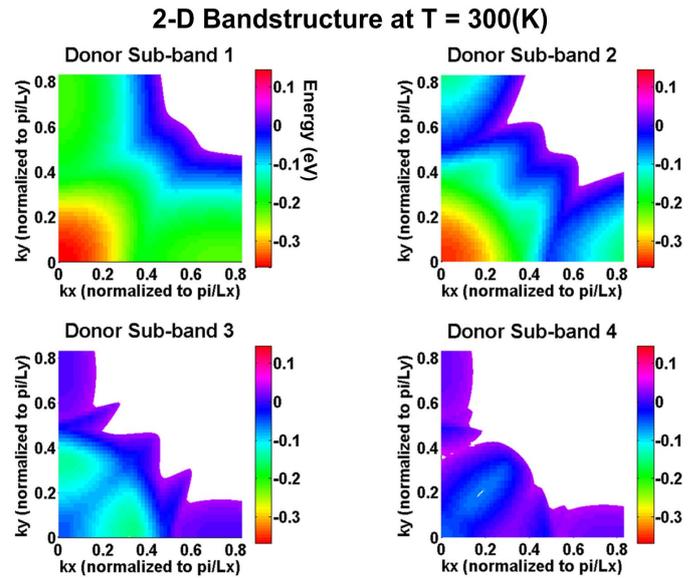

Figure 6. Pseudo-color plots for the first 4 donor sub-bands of the Si:P quantum well 2-D band-structure.

### B. Temperature-dependency of Electronic Properties

Fig. 7 shows the temperature-dependence of the Si:P QW band-structure, where the relative changes of the donor band minimum (DBM) and corresponding Fermi-level position with respect to the result at room temperature are plotted. The result shows that the DBM increases with decreasing temperature. While the Fermi-level position changes with a similar pattern, the Fermi-level change with respect to the DBM turns out to be much smaller. This result indicates that the Fermi-level is strongly pinned near DBM because most of the electrons in the Si:P QW are coming from donor sub-bands. As temperature increases, the electrons start to occupy higher sub-bands, which are less confined by the impurity ion induced potential. Local electrons are more spread inside the QW structure and the number of electrons bound at impurity positions, reduces since the total amount of electrons should be always same as the number of phosphorus ions to satisfy the charge neutrality.

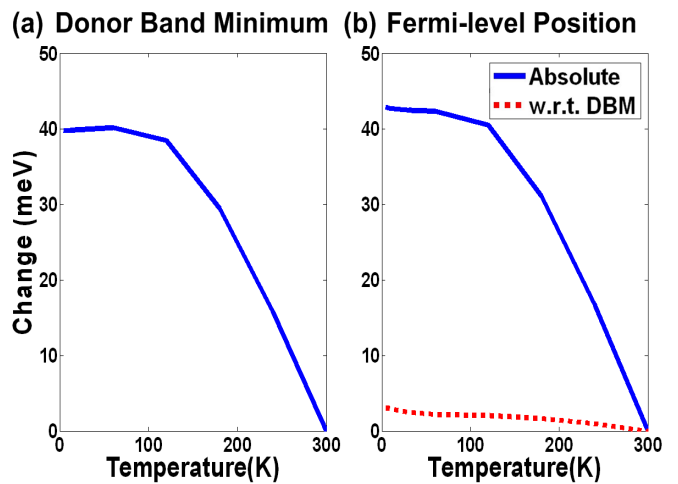

Figure 7. (a) Relative change of the donor band minimum and (b) Fermi-level position with respect to room-temperature.

The temperature-dependence of the potential minimum at impurity positions, which is plotted in Fig. 8, also explains why the impurity ions have fewer bound electrons with increasing temperature since the decreasing potential minimum indicates the impurity ions are less screened losing the bound electrons.

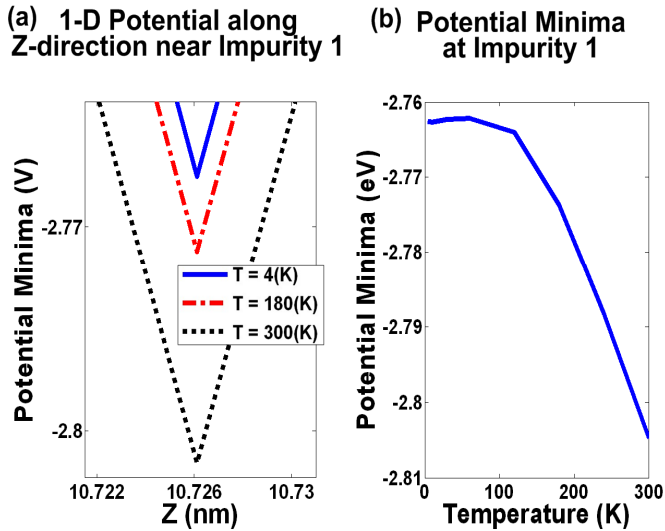

Figure 8. (a) 1-D cut of potential profile along Z-direction near the Impurity 1 and (b) Temperature-dependency of potential minimum at Impurity 1

### IV. CONCLUSION

Temperature-dependent electronic properties of phosphorus δ-doped Si devices in equilibrium have been studied with Si:P quantum wells. For realistic modeling, the spds* tight binding band model has been used with an atomistic representation of impurity ions. Charge and potential self-consistent calculation has been performed with a strongly parallelized Schrödinger-Poisson solver. Further work including exchange corrections and dopant-disorder are currently in progress.

### ACKNOWLEDGMENT

This work has been financially supported by the National Science Foundation (NSF), the Purdue Research Foundation, the Army Research Office and the Semiconductor Research Corporation. NSF-funded computational resources on nanoHUB.org have been extensively used throughout this work. Fruitful discussions with Dr. Mathieu Luisier and the Michele Simmons group in Australia are gratefully acknowledged.